\documentclass{sig-alternate}
\usepackage{amsmath,amssymb,graphicx,epsf,psfrag,epsfig,cite,color}

\def\scalefig#1{\epsfxsize #1\textwidth}

\newtheorem{property}{\bf Property}
\newtheorem{theorem}{\bf Theorem}
\newtheorem{lemma}{\bf Lemma}

\begin{document}

\title{Minimum Information Dominating Set for \\Opinion Sampling}

\numberofauthors{3} 
\author{
\alignauthor
Jianhang Gao\\
       \affaddr{University of California Davis}\\
       \affaddr{Davis, CA, USA}\\
       \email{jhgao@ucdavis.edu}
\alignauthor
Qing Zhao\\
       \affaddr{University of California Davis}\\
       \affaddr{Davis, CA, USA}\\
       \email{qzhao@ucdavis.edu}
\alignauthor 
Ananthram Swami\\
       \affaddr{Army Research Laboratory}\\
       \affaddr{Adelphi, MD, USA}\\
       \email{a.swami@ieee.org}
\and  
}
\maketitle
\begin{abstract}
We consider the problem of inferring the opinions of a social network through strategically sampling a minimum subset of nodes by exploiting correlations in node opinions. We first introduce the concept of information dominating set (IDS). A subset of nodes in a given network is an IDS if knowing the opinions of nodes in this subset is sufficient to infer the opinion of the entire network. We focus on two fundamental algorithmic problems: (i) given a subset of the network, how to determine whether it is an IDS; (ii) how to construct a minimum IDS. Assuming binary opinions and the local majority rule for opinion correlation, we show that the first problem is co-NP-complete and the second problem is NP-hard in general networks. We then focus on networks with special structures, in particular, acyclic networks. We show that in acyclic networks, both problems admit linear-complexity solutions by establishing a connection between the IDS problems and the vertex cover problem. Our technique for establishing the hardness of the IDS problems is based on a novel graph transformation that transforms the IDS problems in a general network to that in an odd-degree network. This graph transformation technique not only gives an approximation algorithm to the IDS problems, but also provides a useful tool for general studies related to the local majority rule. Besides opinion sampling for applications such as political polling and market survey, the concept of IDS and the results obtained in this paper also find applications in data compression and identifying critical nodes in information networks. 
\end{abstract}


\section{Introduction}
\subsection{Information Dominating Set}
In social and information networks, it is often necessary to gauge the general opinion of a large population on a certain issue. Common examples include political polling and market survey of commercial products. Since polling the opinion of a node in the network often incurs a cost (either monetary or in terms of delay), an important question is how to infer the opinion of the entire network through strategically sampling a minimum subset of nodes by exploiting correlations in node opinions. 

This paper presents an algorithmic study of the strategic opinion sampling problem. We first introduce the concept of information dominating set (IDS). A subset of nodes in a given network is an IDS if knowing the opinions of nodes in this subset is sufficient to infer the opinions of the entire network. We focus on two fundamental questions: (i) given a subset of the network, how to determine whether it is an IDS; (ii) how to construct a minimum IDS (i.e., an IDS consisting of a minimum number of nodes) for a given network. The former is referred to as the IDS checker (IDSC) problem, and the latter the minimum IDS (MIDS) problem. 

While the concept of IDS applies to general opinion and opinion correlation models, in this paper, we focus on binary opinions and adopt the local majority rule to model opinion correlation. Specifically, each node in the network has a binary opinion that is consistent with the majority opinion of its neighbors. Local majority rule is commonly used in studying opinion dynamics in social networks (see, for example, \cite{Plott1967, Mustafa&Pekec2001}). Under the local majority model, we show that for a general network, the IDSC problem is co-NP-complete and the MIDS problem is NP-hard. We then focus on networks with special structures, in particular, acyclic networks. We show that in acyclic networks, both IDSC and MIDS problems admit linear-complexity solutions by establishing a connection between the IDS problem and the vertex cover problem.  Our technique for establishing the hardness of the IDS problems is based on a novel graph transformation that transforms the IDS problems in a general network to that in an odd-degree network. This graph transformation technique not only gives an approximation algorithm to our problem, but also provides a useful tool for general studies related to the local majority rule. Furthermore, as a by-product of our complexity analysis, we show that it is NP-complete to determine whether a network can be partitioned into two strong communities with the same size. This result may have implications in the general studies of community structures in social networks.  

Besides the applications in strategic opinion sampling for political polling and market survey, the concept of IDS and the results obtained in this paper also bear significance in identifying critical nodes in information networks. Identifying such critical nodes has important applications in learning and inference under resource constraints as well as security considerations in terms of protecting critical information hubs.  The concept of information dominating set may also be used in data compression, given that an IDS completely represents the information of the entire network.

\subsection{Related Work}\label{sec:related}

Statistical sampling is a classic problem pinioned by Neyman in 1934 \cite{Neyman1934}. Different from the deterministic model and the algorithmic approach taken in this paper, statistical sampling assumes that the value associated with each node a random variable obeying a known probability distribution and designing the sampling strategy amounts to choosing a probability with which each node will be sampled. More recent work on statistical sampling can be found in \cite{Martin&Frankel1987,Heckathom1997,Thompson&Seber1996,Lavallee2007}. 


There are several classic algorithmic problems that are related to the IDS problem. The vertex cover (VC) asks for a (minimum) subset of vertices in a graph such that each edge is adjacent to at least one vertex in this subset. It was proven to be NP-complete by Karp \cite{Karp1972}. Approximation algorithms with a near constant approximation ratio of $2-\Theta(\frac{1}{\sqrt{\log n}})$ were developed in \cite{Karakostas}. Another related algorithmic problem is the dominating set (DS) problem which asks for a subset of vertices such that each vertex in a given graph is either in this set or adjacent to a vertex in this set. The DS problem is also NP-complete \cite{Garey&Johnson1979} and can be approximated within $1-o(1)\log n$ \cite{Feige1998}. The IDS problem studied in this paper is inherently more complex than VC and DS. For instance, as shown in this paper, it is co-NP-complete to verify whether a given subset is an IDS, while VC and DS have trivial polynomial time checker simply based on their definitions. Further discussions on the connections and differences of IDS to VC and DS are given in Sec.~\ref{sec:vcds}.

The local majority rule has been commonly adopted in studying opinion dynamics in social networks (see, for example, \cite{Plott1967, Mustafa&Pekec2001}). The focus of this line of work is on characterizing the evolution of network opinions when each node dynamically changes its opinion by following the majority opinion of its neighbors. The objective of this paper is different: we aim to infer the network opinions \emph{after} the opinion of each node has reached an equilibrium value. 

\section{Problem Formulation}\label{sec:fomu}
\subsection{Information Dominating Set}
\begin{figure}[htbp]
\centering
\scalefig{0.25}\epsfbox{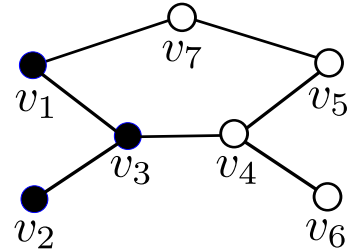}
\caption{The colors of vertices represents their opinions. In this example, the opinion profile is (1,1,1,0,0,0,0) and it is a valid opinion profile. Though the neighbors of both $v_1$ and $v_7$ are half black half white, they are still valid based on the definition.}
\label{fig:validassignment}
\end{figure}

Given a graph $G=(V,E)$ with $n=|V|$ vertices, a \emph{binary opinion profile} $\mathbf{\mu}$ on $G$ is a binary vector $(\mu_{v_1},\ldots,\mu_{v_n})$ indicating where $\mu_{v_i} \in \{0,1\}$ represents the opinion of vertex $v_i$. For a given a binary opinion profile $\mathbf{\mu}$ on $G$, the neighbors of a vertex $v_i$ are partitioned into two groups: the same-minded and opposite-minded neighbors, depending on whether they share the same opinion with $v_i$.  In Fig.~\ref{fig:validassignment}, the same-minded neighbors of $v_3$ are $v_1,v_2$ while its opposite-minded neighbor is $v_4$.

A \emph{valid opinion profile} $\mathbf{\mu}$ under the \emph{local majority rule} in $G$ is a binary opinion profile such that for each vertex $v_i$, the number of its same-minded neighbors is greater than or equal to the number of its opposite-minded neighbors. In other words, the opinion of each vertex is consistent with the majority opinion among its neighbors. And if there is no such majority opinion, this vertex may take any opinion. Fig~\ref{fig:validassignment} demonstrates a valid opinion profile $u$ over the graph. 

The \emph{valid opinion profile set} $U$ of a given graph $G$ is the set of all valid opinion profiles on $G$.

An \emph{information dominating set} (IDS) in a given graph $G$ is a subset of vertices $D\subset V$ such that under any opinion profile, the opinions of vertices in $D$ is sufficient to infer the opinions of all the other vertices. Based on the definition, IDS has an important property as follows. 

\begin{property}\label{pro:ids}
A subset of vertices $D$ in a graph $G$ is an IDS if and only if for any pair of different valid opinion profiles $\mathbf{\mu},\mathbf{\nu}$, there exists a vertex $v\in D$ such that $\mu_v\neq \nu_v$.
\end{property}

\begin{figure}[htbp]
\centering
\scalefig{0.4}\epsfbox{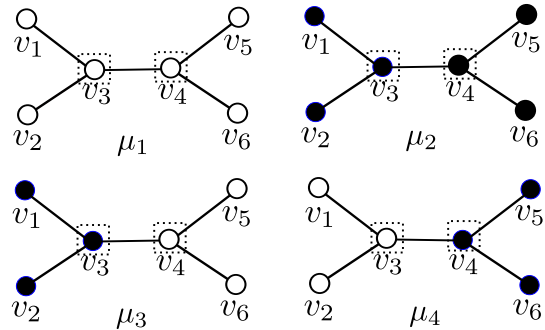}
\caption{There are only four valid opinion profiles on this graph. By Property~\ref{pro:ids}, subset $\{v_3,v_4\}$ is an IDS. }
\label{fig:ids}
\end{figure}

The significance of Property~\ref{pro:ids} is that it provides a way to determine whether a subset of vertices is an IDS or not without considering any specific inferring method. It is used repeatedly in this paper. Fig.~\ref{fig:ids} demonstrates the valid opinion profile set $U$ on a graph and an IDS $\{v_3,v_4\}$.

In this paper, we focus on two problems on IDS. The first problem,  referred to as the IDS checker (IDSC) problem, is to determine whether a given set is an IDS. The second problem we consider is the main objective of this paper, which is to find the minimum IDS (MIDS). In hardness analysis, the corresponding decision problem is: given a graph $G$ and a parameter $k$, whether there exists an IDS $D$ in $G$ with size at most $k$. 

\subsection{Connections to Vertex Cover and Dominating Set Probelms}\label{sec:vcds}
A vertex cover in a graph $G$ is a subset of vertices such that each edge is adjacent to at least one vertex in this subset, equivalently, any vertex in $G$ is either in this subset or all its neighbors are in this subset. It is easy to see that when the opinion correlation model is such that the opinion of one vertex can be completely determined by the opinions of all its neighbors, then a VC of $G$ is an IDS of $G$. Under the local majority rule, when a vertex has an even number of neighbors, its opinion may not be determinable even if the opinions of all of its neighbors are known. Hence a VC is not an IDS in general graphs. In the next section, we will propose an odd-degree graph transformation such that both IDSC and MIDS problem in an arbitrary graph can be solved in an odd-degree graph as the result of the transformation. In the derived odd-degree graph, a VC is an IDS. However, even in odd-degree graphs, an IDS may not be a VC, hence the minimum IDS could be smaller than the minimum VC.

A dominating set is a subset such that any vertex in $G$ is either in this subset or at least one of its neighbor is in the subset. It is can be seen that when the opinion correlation model is such that the opinion of one vertex can be completely determined by one of its neighbors, then a DS of $G$ is an IDS of $G$. 

\begin{figure}[htbp]
\centering
\scalefig{0.35}\epsfbox{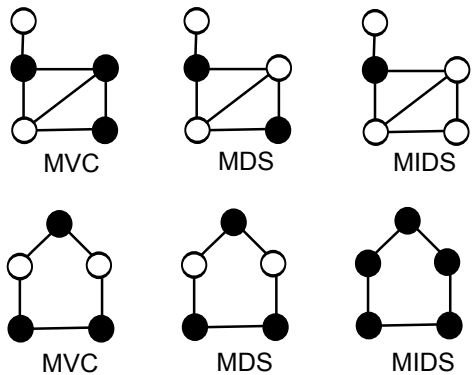}
\caption{In the upper graph, the minimum IDS is smaller than both the minimum VC (MVC) and the minimum DS (MDS). In the lower graph, the minimum IDS (MIDS) contains all the vertices. This is because even if we choose four vertices (without loss of generality, assume we pick the bottom four vertices), the opinion of the remaining vertices cannot be determined if the two vertices on the left side have opinion $0$ and those on the right have opinion $1$.}
\label{fig:ids_vs_vc_ds2}
\end{figure}

As discussed above, under the local majority rule considered in this paper, the opinion of a vertex cannot be completely determined by the opinions of all its neighbors. Hence neither a vertex cover nor a dominating set is an IDS under the local majority rule, and vice versa. Consequently, the size of the minimum VC or the minimum DS in a graph has no direct relationship to the size of the minimum IDS in general graphs. Specifically, the size of the minimum IDS could be larger than the minimum VC or smaller than the minimum DS (Note that a vertex cover is always a dominating set, hence the size of the minimum VC is no less than that of the minimum DS). Fig.~\ref{fig:ids_vs_vc_ds2} demonstrates two examples illustrating the above statement.

Furthermore, the minimum IDS problem is fundamentally different from both problems. Based on the definition, one can easily check whether a given subset is a VC or DS in polynomial time, while in this paper, we show that to determine whether or not a give set is an IDS is co-NP-complete. This imposes difficulties on constructing approximation algorithms because most approximation techniques require a polynomial verifier for the problem. 

\section{Odd-degree Graph Transformation}
By the definition of valid opinion profile under the local majority rule, we may not be able to determine a node's opinion, even if we know the opinions of all its neighbors. Such a case occurs when a vertex has the same number of same-minded and opposite-minded neighbors. For example, vertex $v_7$ in Fig.~\ref{fig:validassignment} can have an opinion of either $0$ or $1$. This imposes difficulties on both the hardness analysis and algorithm design. However, this uncertainty of opinion only occurs if the vertex has an even number of neighbors. In an odd-degree graph where every node has an odd number of neighbors, every vertex will have a unique majority opinion among its neighbors, hence its own opinion can be determined if the opinions of all of its neighbors are known. It thus follows that a vertex cover is an IDS in the odd-degree graph. 

In this section, we propose a way to transform an arbitrary graph $G$ to an odd-degree graph $G'$ such that both the IDSC and MIDS problem in $G$ can be solved by considering $G'$.

\begin{figure}[htbp]
\centering
\scalefig{0.4}\epsfbox{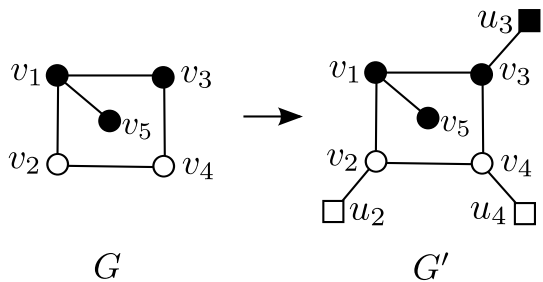}
\caption{An example of the odd-degree transformation from $G$ to $G'$. The round vertices in $G'$ are derived from $G$ and the square vertices are the auxiliary vertices. It also shows the odd-degree transformation from $\mu$ to $\mu'$.}
\label{fig:oddtrans}
\end{figure}

Given an arbitrary graph $G=(V,E)$, we first copy every vertex and edge to $G'$. Then, for every even degree vertex $v_i$ in $G'$, we attach an auxiliary neighbor $u_i$ (see Fig.~\ref{fig:oddtrans}). We call $G'$ the \emph{odd-degree transformation} of $G$. Given any valid opinion profile $\mu$ in $G$, we construct its odd-degree transformation opinion profile $\mu'$ according the following equations:
$$
\begin{cases}
\mu'_{v_i}=\mu_{v_i}\\
\mu'_{u_i}=\mu'_{v_i}.
\end{cases}
$$
In other words, those vertices derived from the original graph take the original opinions, and every auxiliary vertex take the opinion of the vertex it attaches to. Fig.~\ref{fig:oddtrans} demonstrates an example of the odd-degree transformation from $G$ to $G'$ and a valid opinion profile $\mu$ to $\mu'$.

The following two lemmas show that there is bijection between the valid opinion profile sets $U$ in $G$ and $U'$ in $G'$.

\begin{lemma}\label{lma:odt1}
Every valid opinion profile $\mu'$ in $G'$ is an odd-degree transformation of a valid opinion profile $\mu$ in $G$.
\end{lemma}

\begin{lemma}\label{lma:odt2}
There is a valid opinion profile $\mu\in U$ if and only if its odd-degree transformation $\mu'\in U'$ is a valid opinion profile.
\end{lemma}

The above two lemmas establish a bijection between the set of valid opinion profiles in $G$ and $G'$, which serves as a bridge between an IDS in $G$ and that in $G'$. The following theorem establishes a reduction from both IDSC and MIDS in $G$ to those in $G'$.

\begin{theorem}\label{thm:odt}
There exists an IDS $D$ in $G$ if and only if there exists an IDS $D'$ in $G'$ such that for any vertex $v_i\in D$, either $v_i\in D'$ or its auxiliary vertex $u_i\in D'$.
\end{theorem}

Based on Theorem~\ref{thm:odt}, for both the IDSC and MIDS problems, it suffices to consider only odd-degree graphs. Specifically, given a graph $G$, a subset of vertices $D$ is an IDS if $D$ is an IDS in the odd-degree transformation of $G$. And we can find the MIDS $D$ in $G$ by finding the MIDS $D'$ in the odd-degree transformation of $G$ and mapping $D'$ back to $D$ by the procedure in the second part of the above proof. Unless otherwise noted, the graphs considered in the remaining part of this paper are all odd-degree graphs.

\section{IDS Problems in General \\Networks}
\subsection{The IDSC Problem}
In this subsection, we establish the co-NP-completeness of the IDSC problem. To achieve this, we introduce another decision problem in graphs called the \emph{strong community bisection} (SCB) problem. Given a graph with an even number of vertices, the SCB problem asks whether the graph can be partitioned into two strongly connected sub-graphs of equal size, where a sub-graph is called \emph{strongly connected} if the internal degree of every vertex in this sub-graph is strictly greater than its external degree. We first show that the SCB problem is NP-complete even when all the vertices in the given graph have even degree. Then we reduce this NP-complete problem to the IDSC problem.

\subsubsection{NP-completeness of SCB}\label{sec:scb}
The SCB problem is clearly an NP problem. We then focus on reducing a well-known NP-complete problem to the SCB problem in even degree graphs. The problem we are reducing from is the set partition problem (SPP). The SPP asks that whether a set of positive integers $S=\{x_1,\ldots,x_k\}$ can be partitioned into two disjoint subsets $S_1$ and $S_2$ such that the sum of the numbers in $S_1$ equals that of $S_2$. Given such a set of positive integers, we will construct a graph $G$ as follows.

\begin{figure}[htbp]
\centering
\psfrag{a}[c]{$C^1_i$}
\psfrag{b}[c]{$C^2_i$}
\scalefig{0.25}\epsfbox{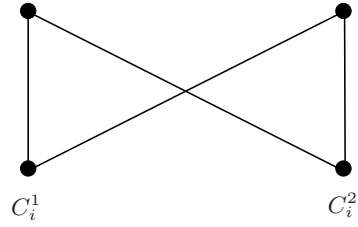}
\caption{An example of the graph component with $x_i=1$. }
\label{fig:twocliques}
\end{figure}

First, for each $x_i\in S$, we construct a sub-graph component with two identical cliques $C^1_i$ and $C^2_i$. The sizes of both cliques are $2x_i$. Then, we connect each vertex in $C^1_i$ to all vertices except its counterpart in $C^2_i$. As a result, an integer $x_i$ is turned into a graph with $4x_i$ vertices and each vertex has $4x_i-2$ neighbors. Fig.~\ref{fig:twocliques} is an example of the graph component corresponding to $x_i=1$.

With all $k$ integers mapping to $k$ connected components, the graph $G$ simply consists of these $k$ disjoint components. Since the component that corresponds to $x_i$ contains $4x_i$ vertices, all with degree $4x_i-2$, graph $G$ is an even degree graph with an even number of vertices ($4\sum_{i=1}^k x_i$). The following theorem establishes the correctness of this reduction.

\begin{theorem}\label{thm:spp2scb}
The set of positive integers $S$ has a equal sum partition if and only if graph $G$ has a strong community bisection.
\end{theorem}

Since the SCB problem is clearly in the NP space and the above reduction can be done in polynomial time, we conclude that the SCB problem  in even degree graphs is NP-complete.

\subsubsection{Co-NP-completeness of IDSC}\label{sec:scb2dids}
The definition of IDS does not imply that IDSC is an NP problem. However, a subset of vertices $D\subset V$ is not an IDS if and only if there exists a pair of different valid opinion profiles $\mu,\nu$ such that the opinion profile on the subset $D$ are identical. Hence it only takes polynomial amount of time to verify whether $D$ is not an IDS. Therefore, IDSC is a co-NP problem. Next, we will reduce SCB in even degree graphs to IDSC.

\begin{figure}[htbp]
\centering
\psfrag{a}[c]{$G_1$}
\psfrag{b}[c]{$G_2$}
\psfrag{c}[c]{$v_1$}
\psfrag{d}[c]{$v_2$}
\scalefig{0.35}\epsfbox{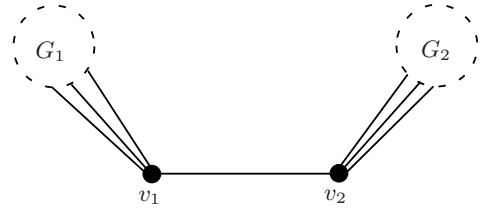}
\caption{The parts $G_1$ and $G_2$ are copies of $G$. All vertices in $G_1$ and $G_2$ are connected to $v_1$ and $v_2$, respectively. Additionally, vertices $v_1$ and $v_2$ are connected.}
\label{fig:spp2dids}
\end{figure}

Given a graph $G$ where each vertex has even number of neighbors, we construct a new graph $G'$ as follows. We first make two copies of $G$: $G_1$ and $G_2$. Next, we add two additional vertices, $v_1$ and $v_2$. Finally, we connect all vertices in $G_1$ to $v_1$, all vertices in $G_2$ to $v_2$ and $v_1$ to $v_2$. Fig.~\ref{fig:spp2dids} demonstrates the structure of $G'$.

The following theorem establishes the co-NP-hardness of the IDSC problem even when the subset contains all but two connected vertices in the graph.
\begin{theorem}\label{thm:scb2dids}
Let $D$ denote the set of all vertices in $G'$ except $v_1$ and $v_2$. Given an even degree graph $G$, it has a strong community bisection if and only if the $D$ is not an IDS in $G'$.
\end{theorem}

\subsection{The MIDS Problem}
Now we consider the problem of MIDS. Since IDSC is co-NP-complete, unless P equals NP, we cannot conclude that MIDS is in the NP space. In this section, we prove that the MIDS problem is NP-hard,  not necessarily NP-complete if MIDS does not belong to the NP space.

The construction of the reduction is a combination of the previous two reductions. More specifically, we reduce the SPP problem with an integer set $S$ to the MIDS problem in a graph $G'$ in two steps. First, we construct a graph $G$ based on the given integer set $S$ by the same procedure in the reduction in Section~\ref{sec:scb}. Second, we construct a new graph $G'$ from $G$ by following the procedure in Section~\ref{sec:scb2dids}. The following theorem establishes the reduction from SPP to MIDS.

\begin{theorem}\label{thm:spp2mids}
Let $S$ be a set containing $k$ positive integers. It has a equal sum partition if and only if there does not exist an IDS in $G'$ with size at most $2k$.
\end{theorem}

\section{IDS Problems In Acyclic\\ Networks}
In this section, we consider both IDSC and MIDS problem on acyclic networks. An acyclic network is a forest (i.e., a collection of trees). Since each connected component of the network can be considered separately when studying the IDS problems, it suffices to focus on trees. We show, in Lemma~\ref{lma:vc2ids}, that an IDS without any leaf node is a vertex cover in an odd-degree tree. Since both an IDS or a vertex cover with leaf vertex can be transformed into a same size IDS or a vertex cover without any leaf vertex, respectively, we can solve IDSC and MIDS by solving the vertex cover problem
\begin{lemma}
\label{lma:vc2ids}
Given an odd-degree tree $G$, an IDS that does not contain any leaf is also a vertex cover in $G$.
\end{lemma}

Lemma~\ref{lma:vc2ids} only considers an IDS without any leaf. The following lemma extends this result to any IDS.

\begin{lemma}
\label{lma:midsnoleaf}
Given any IDS $D$, there exists an \mbox{IDS} $D'$ that contains no leaf nodes and has a size smaller than or equal to $D$.
\end{lemma}

With Lemma~\ref{lma:midsnoleaf}, we can solve the IDSC on a tree by checking whether its non-leaf transformation is a vertex cover. Furthermore, the following theorem provide us a way to find the MIDS.

\begin{theorem}
The non-leaf minimum vertex cover is a minimum IDS.
\end{theorem}

Since the non-leaf minimum vertex cover can be solved in linear time by a greedy algorithm, we can solve the MIDS on trees in linear time.

\section{Conclusion}
In this paper, we introduce the concept of information dominating set (IDS) for strategic opinion sampling in social networks and identifying critical nodes in information networks. Based on a novel odd-degree graph transformation, we show that it is enough to consider the problem only in odd-degree graphs. We establish the NP-hardness of both the problem of finding the minimum IDS and the problem of determining whether a given subset is an IDS. We further consider both problems in acyclic networks and developed linear time complexity solutions. This graph transformation technique provides a useful tool for general studies related to the local majority rule. Furthermore, as a by-product of our complexity analysis, we show that it is NP-complete to determine whether a network can be partitioned into two strong communities with the same size. This result may have implications in the general studies of community structures in social networks. Besides opinion sampling for applications such as political polling and market survey, the concept of IDS and the results obtained in this paper also find applications in data compression and identifying critical nodes in information networks. 


\bibliographystyle{ieeetr}

{\footnotesize
\begin{thebibliography}{10}
\vspace{1em}

\bibitem{Plott1967}
C. R. Plott, ``A Notion of Equilibrium and Its Possibility Under Majority Rule'', In American Economic Review, volume 57, 1967.

\bibitem{Mustafa&Pekec2001}
N.~Mustafa, A.~Pekec, ``Majority Consensus and the Local Majority Rule'',Automata, Languages and Programming Lecture Notes in Computer Science Volume 2076, 2001, pp 530-542.



\bibitem{Neyman1934}
J.~Neyman, ``On the Two Different Aspects of the Representative Method: The Method of Stratified Sampling and the Method of Purposive Selection'' Journal of the Royal Statistical, 1934,
Society 97:558-625.

\bibitem{Martin&Frankel1987}
F.~Martin, L.~Frankel,``Fifty Years of Survey Sampling in the United States'', Public Opinion Quarterly 51(Part 2), 1987, S127-38.


\bibitem{Heckathom1997}
D.~Heckathorn, ``Respondent-Driven Sampling: A New Approach to the Study of Hidden Populations'', Social Problems, 1997, vol. 44, No. 2, pp. 174-199.

\bibitem{Thompson&Seber1996}
S.~Thompson, G.~Seber, \emph{Adaptive Sampling}, Wiley,  1996.


\bibitem{Lavallee2007}
P.~Lavallece, \emph{Indirect Sampling}, Springer, 2007.





\bibitem{Karp1972}
R.M.~Karp, ``Reducibility Among Combinatorial Problems'', Complexity of Computer
Computations, Plenum Press, 1972, pp. 85-103.


\bibitem{Karakostas}
G.~Karakostas,``A better approximation ratio for the Vertex Cover problem'',  Automata, Languages and Programming
Lecture Notes in Computer Science Volume 3580, 2005, pp. 1043-1050

\bibitem{Garey&Johnson1979}
M.R.~Garey, D.S.~Johnson \emph{Computers and Intractability: A guide to the theory of NP-completeness,} Freeman, San Francisco, 1978.

\bibitem{Feige1998}
U.~Feige, ``A Threshold of ln n for Approximating Set Cover'', Journal of the ACM, Vol. 45, No. 4, July 1998, pp. 634-652.






\end{thebibliography}
}

\end{document}